\documentclass[reprint,
superscriptaddress,
amsmath,amssymb,
aps,prd,floatfix,
twocolumn,
showpacs,
]{revtex4-2}

\usepackage{amsthm}
\usepackage{tikz}
\usepackage{graphicx}
\usepackage{dcolumn}
\usepackage{bm}
\usepackage{braket}
\usepackage[english]{babel}
\usepackage{hyperref}
\usepackage{amsmath}
\usepackage{graphicx}
\usepackage{float}
\usepackage{siunitx}
\usepackage{xcolor}
\usepackage{xspace}
\usepackage{eucal}
\usepackage{ulem}
\usepackage{multirow}
\usepackage[utf8]{inputenc}
\usepackage{pgfplots}
\pgfplotsset{compat=1.14}
\usepackage{upgreek}
\usepackage{siunitx}
\usepackage{gensymb}
\usepackage{amsmath}
\usepackage[justification=justified,singlelinecheck=false]{caption}
\usepackage{subcaption}
\usepackage{textcomp}
\usepackage{comment}
\usepackage{cleveref}

\Crefname{figure}{Figure}{Figures}
\usepackage{xcolor,colortbl}
\usepackage{todonotes}
\usepackage{tabularx}
\usepackage{booktabs}
\usepackage{comment}
\DeclareSIUnit\g{g}
\DeclareSIUnit\gal{Gal}
\DeclareSIUnit\torr{Torr}
\DeclareSIUnit\bar{Bar}
\DeclareSIUnit\kelvin{K}
\DeclareSIUnit\inch{inch}
\DeclareSIUnit\joule{J}
\DeclareSIUnit\rad{rad}

\begin{document}

\title{A characterization method for low-frequency environmental noise in LIGO}
\author{Guillermo Valdes}\email[Electronic mail: ]{gvaldes@tamu.edu}
\author{Adam Hines}
\author{Andrea Nelson}
\author{Yanqi Zhang}
\author{Felipe Guzman}
\affiliation{Texas A\&M University, Aerospace Engineering \& Physics, College Station, TX 77843}%


\begin{abstract}
We present a method to characterize the noise in ground-based gravitational-wave observatories such as the Laser Gravitational-Wave Observatory (LIGO).
This method uses linear regression algorithms such as the least absolute shrinkage and selection operator (LASSO) to identify noise sources and analyzes the detector output versus noise witness sensors to quantify the coupling of such noise.
Our method can be implemented with currently available resources at LIGO, which avoids extra coding or direct experimentation at the LIGO sites.
We present two examples to validate and estimate the coupling of elevated ground motion at frequencies below 10 Hz with noise in the detector output.
\end{abstract}

\maketitle
\label{sec:Intro}
Ground-based gravitational-wave observatories, such as LIGO \cite{TheLIGOScientific:2014jea}, Virgo \cite{VIRGO:2014yos}, and KAGRA \cite{KAGRA:2018plz}, are sensitive instruments capable of measuring strain changes in the order of $10^{-23}$.
The gravitational-wave detectors of these advanced observatories consist of a  L-shaped laser interferometer and state-of-the-art systems to isolate their components from environmental noises.
Still, the observatories are susceptible to disturbances, such as earthquakes, thunderstorms, and human activity \cite{LIGO:2021ppb}.
These disturbances couple into the detectors' output, causing noise such as scattered-light noise (or scattering) \cite{Accadia:2010zzb} at different frequencies.
Therefore, it is necessary to identify and quantify the contribution of external disturbances to noise in the output to mitigate it \cite{Davis:2022dnd}.

The complex design and constant upgrading of these detectors make it extremely hard to find the noise coupling mechanism, which can be linear or nonlinear, even with the thousands of diagnostic sensors monitoring their status. 
We can simplify this task by creating adaptive tools, such as \textit{GWadaptive\_scattering} \cite{Bianchi:2021unp, Longo:2021avq}, that use the data recorded by installed sensors and computational methods, allowing for long-duration studies in observation runs.

To identify and quantify the contribution of environmental disturbances to the noise in  LIGO,  scientists have performed hardware excitations (so-called injections) at targeted locations to study how ground motion couples into the detector \cite{AdvLIGO:2021oxw}. This task is time-consuming and challenging to perform often and in many areas. 
Algorithms using machine learning have been developed to identify the contribution of instrumental and external disturbances to the noise in LIGO \cite{Walker_2018}. 
For example, the least absolute shrinkage and selection operator (LASSO) regression \cite{LASSO_paper_1996} has been used as a feature selector to identify from a list of signals recorded by diagnostic sensors which one contributes the most to the behavior of the \textit{binary neutron star inspiral range} (a figure of merit quantifying the observatory performance \cite{Allen:1999yt}). 
That application does not investigate detector noise data at different frequencies and implements limited pre-processing of the data.  


Here, we present a computational method to characterize the contribution of ground motion to the gravitational-wave detectors noise at distinct frequencies using archived diagnostic data, different signal pre-processing, and machine learning algorithms. 
Our method avoids direct experiments at the detectors' sites, leading to more frequent and longer-term studies.

In this article, we describe how the injections have been performed and contributions are quantified by calculating coupling factors. 
Then, we discuss the LASSO algorithm and how we use it to identify the highest contributions from ground motion to the detector noise. 
Finally, we present two examples to validate the performance of our tool to characterize the LIGO Livingston detector noise.  
In this article,we refer to gravitational-wave detectors as simply \textit{detectors}, and to the data monitoring gravitational-wave signals as \textit{detector output}.

\label{sec:methods}
Injections are completed by targetly perturbing different areas in the detectors. 
This is achieved by attaching electronic shakers to distinct locations of the vacuum chambers and beam tubes enclosing the optics and laser.
Then, we compare the amplitude spectral densities (ASDs) of the detector output and the witness sensors during the injection to the ASDs calculated when both are at background noise levels, e.g., when the detector is ready to observe and no disturbances are present. 
To complete that comparison, we estimate the \textit{coupling function} CF at some frequency $f$, given by the following equation:

\begin{equation}
\text{CF}(f) = \sqrt{\frac{[Y_{\text{inj}}(f)]^2-[Y_{\text{bkg}}(f)]^2}{[X_{\text{inj}}(f)]^2-[X_{\text{bkg}}(f)]^2}}, 
\label{eq1}
\end{equation}

where $X_{\text{bkg}}$ and $X_{\text{inj}}$ are the ASDs of the witness sensor at background and injection times, and $Y_{\text{bkg}}$ and $Y_{\text{inj}}$ are the ASDs of detector output at background and injection times.

We refer to the value of a coupling function at a single frequency bin $f_\text{bin}$ as \textit{coupling factor}.
We can calculate the coupling factor with a similar method. 
First, we filter the signal for the desired frequency bin, and then we calculate the root-mean-square (RMS) value.
Finally, we use Equation \ref{eq2} to estimate the coupling factor:

\begin{equation}
\text{CF}(f_\text{bin}) =\frac{y_{\text{inj}} - y_{\text{bkg}}}{x_{\text{inj}} - x_{\text{bkg}}},
\label{eq2}
\end{equation}

where $x_{\text{bkg}}$ and $y_{\text{bkg}}$ are the mean value of the band-limited RMS signals of the witness sensor and detector output at background times, and $x_{\text{inj}}$ and $y_{\text{inj}}$ are the maximum value of the band-limited RMS signals of the witness sensor and detector output at injection times.

We use LASSO to find the contribution of the ground motion at different frequencies to the noise in the detector output. 
Here, LASSO will try to reconstruct a target signal (detector output) with a set of input signals (ground motion) and assign coefficients to each input. 
The higher the coefficient the higher the contribution. 
A negative coefficient means that the input signal is negatively related to the target. 
We ignore negative coefficients since those do not have physical meaning for our purposes. 
The LASSO algorithm uses a parameter \textit{alpha}. 
When alpha is 0, LASSO regression produces the same coefficients as a linear regression. 
When alpha is large, all coefficients are zero.

The proposed method consists of two parts: feature selection and quantification.
We use feature selection to identify the source of the noise, for which we use LASSO.
We quantify the contribution of these noise sources to the noise calculating the coupling factors,  similar to what LIGO scientists do when performing injections.
Additionally, we estimate the contribution by analyzing the scatter plot of the detector output versus ground motion.

\label{sec:results}
Here we test our method, using two examples when injections were performed to show the feature selection work and to estimate the coupling factor using the previous method and our method. 

\begin{figure}
\includegraphics[width=0.94\columnwidth]{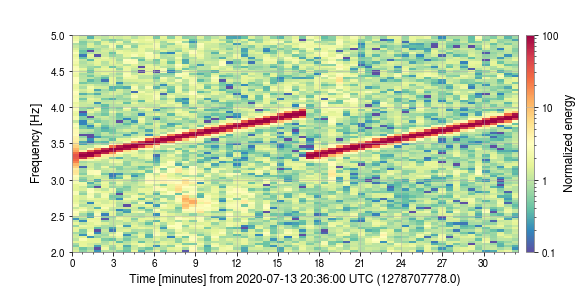}
\caption{\label{fig:ex1a} Spectrogram of a witness sensor (accelerometer) at LIGO Livingston during an injection, where an electrodynamic shaker is attached to the vacuum walls enclosing the gravitational-wave detector and then driven at different frequencies.}
\end{figure}

\begin{figure}
\includegraphics[width=0.94\columnwidth]{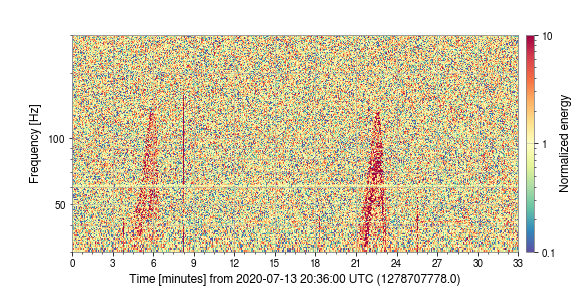}
\caption{\label{fig:ex1b} Spectrogram of the detector output at the LIGO Livingston showing the noise generated by an injection. The detector output and the witness sensor data during the injection are compared to characterize the noise coupling.}
\end{figure}

In this example, we use data when an injection was performed on the beam tube of the Y-end (the top end of the L-shaped interferometer) at LIGO Livingston.
This injection consists of two 16-minute periodic signals with a frequency sweep from 3~\si{\hertz} to 4~\si{\hertz} and witnessed by the accelerometer located on the Y-end beam tube. 
It was found that the excitation at 3.5~\si{\hertz} generated noise in the detector output between 30~\si{\hertz} and 150~\si{\hertz} \cite{alog:effler_injection}. 
We show the spectrograms of the witness accelerometer and the detector output during the injection in Fig.~\ref{fig:ex1a} and Fig.~\ref{fig:ex1b}, respectively. 

We use three input signals to test LASSO: the witness accelerometer signal band-passed between 3.4~\si{\hertz} to 3.6~\si{\hertz}, the signal of an accelerometer located at the X-end (bottom right end of the L-shaped interferometer) band-passed between 3.4~\si{\hertz} to 3.6~\si{\hertz}, and the witness accelerometer signal but band-passed between 6~\si{\hertz} to 9~\si{\hertz}.
We identify these signals as x1, x2, and x3 respectively.
Therefore, we have a signal correlated with the noise in the detector (x1) and two signals that do not contribute, either because they come from an accelerometer far from the injection site (x2) or have a frequency unrelated to the injection (x3).
The target signal is the detector noise band-passed between 70~\si{\hertz} to 90~\si{\hertz} to avoid the power line frequency and its resonances.

In preparation for the application of the LASSO feature selection, we calculate the $\log_{10}$ of the detector noise signals since those reach values of the order of $10^{-23}$.
Then, we calculate the RMS and standardize the input and target signals.
For a set of alpha = [0.05,0.09], the highest LASSO coefficient consistently corresponds to the RMS signal of the witness accelerometer band-passed between 3.4~\si{\hertz} to 3.6~\si{\hertz}. 
The other coefficients are either negative or smaller by two orders of magnitude, as shown in Table \ref{tab:alpha1}.

\begin{table}
\caption{\label{tab:alpha1}LASSO coefficient for the injection example. 
Signal x1 is the witness accelerometer band-passed between 3.4~\si{\hertz} to 3.6~\si{\hertz}, x2 is the signal of an accelerometer located at the X-end band-passed between 3.4~\si{\hertz} to 3.6~\si{\hertz}, and x3 the witness accelerometer signal but band-passed between 6~\si{\hertz} to 9~\si{\hertz}. 
For a set of alpha = [0.05,0.09], the highest LASSO coefficient consistently corresponds to the x1.}
\begin{ruledtabular}
\begin{tabular}{lrrrrr}
alpha & 0.05  & 0.06  & 0.07  & 0.08 & 0.09 \\
\hline
\rowcolor{lightgray}x1    & 0.60  & 0.59  & 0.58  & 0.57 & 0.56 \\
x2    & -0.03 & -0.02 & -0.01 & 0.00 & 0.00 \\
x3    & 0.00  & 0.00  & 0.00  & 0.00 & 0.00\\
\end{tabular}
\end{ruledtabular}
\end{table}

Using Equation \ref{eq2}, we find that the coupling factor of the detector noise frequency bin 70~\si{\hertz} to 90~\si{\hertz} to the Y-end beam tube accelerometer frequency bin 3.4~\si{\hertz} to 3.6~\si{\hertz} is \num{1.4d-24}~\si{ strain\per\nano\meter\per\square\second}.

\begin{figure}[ht]
\includegraphics[width=0.94\columnwidth]{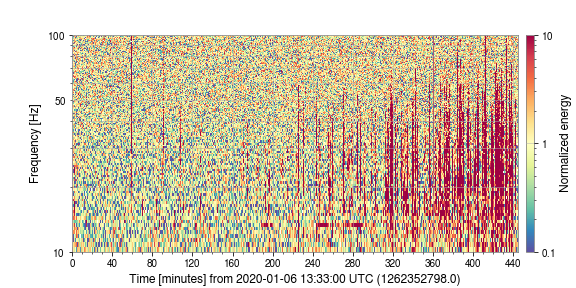}
\caption{\label{fig:ex2} LIGO Livingston spectrogram with elevated noise in the frequency range 10~\si{\hertz} to 100~\si{\hertz}.}
\end{figure}

In this example, we use a 440~\si{\minute} data segment of observing time, when no injections where performed. 
We chose this segment because the first 180~\si{\minute} have low noise in the detector output and then, the noise increases in the last 260~\si{\minute} over frequencies between 10~\si{\hertz} to 100~\si{\hertz}, as shown in the detector output spectrogram in Fig.~\ref{fig:ex2}.

Here, we use the readings from the installed tri-axial seismometers as witness sensors.  
These sensors are located on the ground inside the buildings housing the L-shaped interferometer.
The seismometers data is band-passed at distinct frequency bands and its RMS is calculated and stored in the LIGO diagnostics database.

In our LASSO test, the input signals are the seismometers data listed in Table \ref{tab:list} and the target signal is the RMS of the band-passed detector output  frequencies 20~\si{\hertz} to 40~\si{\hertz} where much of the noise is concentrated. 

\begin{table}
\caption{\label{tab:list} List of LIGO Livingston seismometers channels.  Channels L1:ISI-GND\_STS\_xx\_BLRMS\_yy.mean,m-trend store one sample per minute of band-limited RMS (BLRMS) data, where xx defines the location and degree-of-freedom measured by the seismometer, and yy the frequency range of the band-pass filter employed. 
For example, L1:ISI-GND\_STS\_ETMY\_Y\_BLRMS\_30M\_100M corresponds to readings of the Streckeisen tri-axial seismometer (STS) located on the ground (GND) inside the building enclosing the end test mass Y (ETMY) of the LIGO Livingston (L1) detector, measuring accelerations in the Y-direction filtered for the frequencies 30~\si{\milli\hertz}--100~\si{\milli\hertz}. These seismometers are part of its internal seismic isolation (ISI) system.  We highlight the seismometer with the highest LASSO coefficient obtained in example 2.} 
\centering
\resizebox{\columnwidth}{!}{%
\begin{tabular}{llr}
\hline
\hline
Short&	Channel&	Frequency\\ 
Name&   &Band [\si{\hertz}]	\\
 \hline
A&	L1:ISI-GND\_STS\_ETMY\_Y\_BLRMS\_30M\_100M.mean,m-trend&		0.03-0.1		\\
\rowcolor{lightgray}B&	L1:ISI-GND\_STS\_ETMY\_Y\_BLRMS\_100M\_300M.mean,m-trend&	0.1-0.3			\\
C&	L1:ISI-GND\_STS\_ITMX\_X\_BLRMS\_300M\_1.mean,m-trend&			0.3-1			\\
D&	L1:ISI-GND\_STS\_ETMX\_X\_BLRMS\_1\_3.mean,m-trend&			    1-3				\\
E&	L1:ISI-GND\_STS\_ETMY\_Y\_BLRMS\_1\_3.mean,m-trend&			    1-3				\\
F&	L1:ISI-GND\_STS\_ITMY\_Z\_BLRMS\_1\_3.mean,m-trend&			    1-3				\\
G&	L1:ISI-GND\_STS\_ETMX\_X\_BLRMS\_3\_10.mean,m-trend&		    3-10			\\
H&	L1:ISI-GND\_STS\_ETMY\_Y\_BLRMS\_3\_10.mean,m-trend&		    3-10			\\
I&	L1:ISI-GND\_STS\_ITMY\_Z\_BLRMS\_3\_10.mean,m-trend&		    3-10			\\
\hline
\hline
\end{tabular}%
}
\end{table}

In preparation for the LASSO feature selection application, we smooth the input and target signals by applying a low-pass filter with a cutoff frequency at 277.8~\si{\micro\hertz} ($\sim$1/60~\si{\minute}). 
We also remove the first 60~\si{\minute} and the last 10~\si{\minute} of the data, because we want to remove possible data transients from the analysis right after the detector enters observing mode and before it loses such status. 

Table \ref{tab:alpha2} shows the LASSO coefficients for alpha = [0.07,0.18] and their mean square error (MSE), which is an indicator of their significance.
In this example, the seismometer B consistently shows the highest LASSO coefficient. 
This indicates that for the list of sensors in Table \ref{tab:list}, the ground motion in the frequency band 0.1~\si{\hertz} to 0.3~\si{\hertz}, witnessed by the seismometer in the building enclosing the the end test mass Y, contributes the most to the 20~\si{\hertz} to 40~\si{\hertz} noise in the detector output.

\begin{table}
\caption{\label{tab:alpha2}LASSO coefficient example 2 for distinct alpha values.  
The signal from seismometer B (see Table \ref{tab:list}) consistently shows the highest LASSO coefficient. 
We include the mean square error (MSE) to quantify the significance of the results.}
\begin{ruledtabular}
\begin{tabular}{lrrrrr}
alpha & 0.06 & 0.09 & 0.12 & 0.15 & 0.18 \\
\hline
MSE   & 0.02 & 0.03 & 0.05 & 0.06 & 0.07 \\
\hline
A     & 0.07 & 0.05 & 0.01 & 0.00 & 0.00 \\
\rowcolor{lightgray}
B     & 0.70 & 0.77 & 0.77 & 0.76 & 0.74 \\
C     & 0.00 & 0.00 & 0.00 & 0.00 & 0.00 \\
D     & 0.07 & 0.11 & 0.10 & 0.09 & 0.07 \\
E     & 0.14 & 0.02 & 0.00 & 0.00 & 0.00 \\
F     & 0.00 & 0.00 & 0.00 & 0.00 & 0.00 \\
G     & 0.07 & 0.00 & 0.00 & 0.00 & 0.00 \\
H     & 0.00 & 0.00 & 0.00 & 0.00 & 0.00 \\
I     & 0.00 & 0.00 & 0.00 & 0.00 & 0.00 \\
\end{tabular}
\end{ruledtabular}
\end{table}

\begin{figure}
\includegraphics[width=0.94\columnwidth]{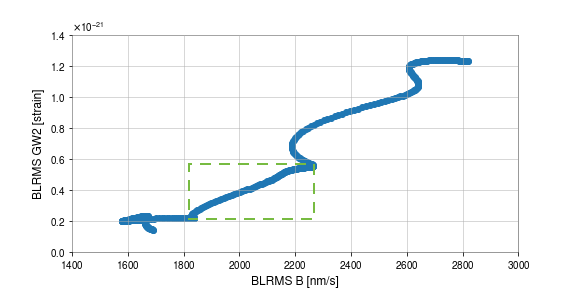}
\caption{\label{fig:svm1} Scatter plot of gw2 (detector noise data) against BLRMS B (ground motion data). We performed a linear fitting to this segment and calculated the slope m = \num{7.47d-25}~\si{ strain\per\nano\meter\per\second} in the region 1850~\si{\nano\meter\per\second} to 2150~\si{\nano\meter\per\second} (dashed rectangle).}
\end{figure}

In Fig. \ref{fig:svm1} we show the scatter plot of these two signals.
Here, we recognize three things: (a) there is no coupling for velocities below 1850~\si{\nano\meter\per\second}, (b) the coupling is linear in the region 1850~\si{\nano\meter\per\second} to 2150~\si{\nano\meter\per\second} (dashed rectangle), and the slope (or coupling factor) is \num{7.47d-25}~\si{ strain\per\nano\meter\per\second}, and (c) the coupling is not longer linear when the ground motion is higher than 2150~\si{\nano\meter\per\second}.

\label{sec:conclusion}
In this paper, we show a computational method to characterize noise in the LIGO observatory output and identify the source of the noise using ground motion sensors. 
Our method showed to be effective using ground motion frequencies below 10~\si{\hertz}. 
It takes advantage of the ground motion noise to help with the characterization procedures, reducing the number of injections necessary to estimate the coupling factors and the inconvenience of finding electrodynamic shakers with operational ranges below 10~\si{\hertz}.  

We validated our method by obtaining the same results as those obtained through a known injection of an intentional disturbance.
In another example, we showed the strong correlation between the noise in the LIGO Livingston gravitational-wave detector at 20~\si{\hertz}--40~\si{\hertz} and the ground motion in the range 0.1~\si{\hertz}--0.3~\si{\hertz}.
We identified three parts in the plot detector output noise versus ground motion. 
The first part showed the minimum value witnessed by the ground motion sensor before the noise is present in the detector output.   
The second part was linear and we used it to calculate the coupling factor. 
The third part indicated the minimum value before the coupling is no longer linear.
The long-term application of this method will be helpful to set upper limits noise contributions and the coupling mechanisms that might be possible to mitigate them.

For our method, we found that a smoothing procedure such as low-pass filtering the signals before using the LASSO algorithm improves the results.
The selection of the right LASSO alpha value remains challenging.
We recommend an iteration over different alpha values and metrics to determine the linear regression effectiveness, such as mean squared error, to automate the selection of alpha. 

Our method intends to work as an automated tool that uses the local environmental noise sources and monitors to characterize the detectors of these gravitational-wave observatories.\\
\newline
The authors have no conflicts to disclose.\\
\newline
G. Valdes led the analysis design, the data collection, and the writing process.
F. Guzman led the funding acquisition. 
A. Hines, A. Nelson, and Y. Zhang contributed to the data collection, analysis implementation, and writing review.
F. Guzman and G. Valdes supervised the overall project. 
All authors approved the final version of the manuscript.\\
\newline
Raw data were generated at the Laser Interferometer Gravitational-Wave Observatory (LIGO). Derived data supporting the findings of this study are available from the corresponding author upon reasonable request.

\subsection*{Acknowledgements}
We acknowledge support from the NSF grant PHY-2045579. 
We also acknowledge the discussions with the members of Detector Characterization Group of the LIGO Scientific Collaboration. 
For this paper, we use the data from the Advanced LIGO detectors and we used the LIGO computing clusters to perform the analysis and calculations.
LIGO was constructed by the California Institute of Technology and Massachusetts Institute of Technology with funding from the National Science Foundation and operates under Cooperative Agreement No. PHY-1764464. Advanced LIGO was built under Grant No. PHY-0823459.

\bibliography{References}

\end{document}